%% file: manuscript_arxiv.tex
\documentclass[a4paper,journal,11pt,draftclsnofoot,onecolumn]{IEEEtran}

\usepackage{dsfont}
\usepackage{amsmath}
\usepackage{amssymb}
\usepackage{amsfonts}
\usepackage{graphicx}
\usepackage{amsmath}
\usepackage[all,poly]{xy}
\usepackage{multirow}
\usepackage{algorithm}
\usepackage{algorithmic}
\usepackage{color}
\usepackage[noadjust]{cite}
\usepackage{caption}
\usepackage{subcaption}
\usepackage{url}
\newcommand{\best}[1]{\textcolor[rgb]{0.00,0.00,1.00}{\mathbf{#1}}}
\newcommand{\second}[1]{\mathbf{#1}}

\newcommand{\figextension}{png}

\input com_notations.tex

\input notations.tex

\title{A comparison of nonlinear mixing models for vegetated areas using simulated and real hyperspectral data}

\author{\vspace{1cm}Nicolas Dobigeon$^{(1)}$, Laurent Tits$^{(2)}$, Ben Somers$^{(3)}$, \\Yoann Altmann$^{(1)}$ and Pol Coppin$^{(2)}$\\
\vspace{0.5cm}
\normalsize $^{(1)}$ University of Toulouse, IRIT/INP-ENSEEIHT/T\'eSA, Toulouse, France\\
\small\texttt{\{Nicolas.Dobigeon,Yoann.Altmann\}@enseeiht.fr}\\
\normalsize $^{(2)}$ Department of Biosystems, Katholieke Universiteit Leuven, 3001 Leuven, Belgium\\
\small\texttt{\{Laurent.Tits,Pol.Coppin\}@biw.kuleuven.be}\\
\normalsize $^{(3)}$ Division Forest, Nature and Landscape, Katholieke Universiteit Leuven, 3001 Leuven, Belgium\\
\small\texttt{ben.somers@ees.kuleuven.be}
\thanks{Part of this work has been funded by the Hypanema ANR Project
n$^\circ$ANR-12-BS03-003 and the Research Foundation Flanders (FWO)
project G.0677.13N.}}

\begin{document}

\maketitle

\begin{abstract}
Spectral unmixing is a crucial processing step when analyzing
hyperspectral data. In such analysis, most of the work in the
literature relies on the widely acknowledged linear mixing model to
describe the observed pixels. Unfortunately, this model has been
shown to be of limited interest for specific scenes, in particular
when acquired over vegetated areas. Consequently, in the past few
years, several nonlinear mixing models have been introduced to take
nonlinear effects into account while performing spectral unmixing. These models have been proposed
empirically, however without any thorough validation. In this paper,
the authors take advantage of two sets of real and physical-based
simulated data to validate the accuracy of various nonlinear models
in vegetated areas. These physics-based models, and
their corresponding unmixing algorithms, are evaluated with respect
to their ability of fitting the measured spectra and of providing an
accurate estimation of the abundance coefficients, considered as the
spatial distribution of the materials in each pixel.
\end{abstract}
%

\section{Introduction}\label{sec:introduction}
Spectral unmixing (SU) of hyperspectral images consists of
extracting the spectral responses $\Vmat{1},\ldots,\Vmat{\nbmat}$ of
the $\nbmat$ macroscopic materials (or \emph{endmembers}) present in
the imaged scene and, for each pixel $\Vpix{\nopix}$ of the image
($\nopix=1,\ldots,\nbpix$), estimating the corresponding proportions
$\abond{1}{\nopix},\ldots,\abond{\nbmat}{\nopix}$ (or
\emph{abundances}) that represent the spatial distributions of these
materials over the area of interest \cite{Bioucas2012jstars}. The
first automated unmixing techniques have been proposed in the early
1990's \cite{Keshava2002}. When no prior knowledge is available
regarding the studied scene, SU can be usually decomposed into two
successive steps. First, the endmembers are extracted from the image
and, subsequently, the proportions of the materials are estimated in
a so-called \emph{inversion} step. A vast majority of the endmember
extraction algorithms (EEA) and inversion techniques exploit some
geometrical concepts that are intrinsically related to an assumption
of a linear mixing process to explain the observed pixels. In other
words, under this linear mixing model (LMM), each observed pixel of
a given image is assumed to result from the linear combination of
the $\nbmat$ endmember spectra
\begin{equation}
  \label{eq:LMM}
  \Vpix{\nopix}^{(\textrm{LMM})} = \sum_{\nomat=1}^{\nbmat} \abond{\nomat}{\nopix}
  \Vmat{\nomat} + \Vnoise{\nopix} = \MATmat\Vabond{\nopix} +
  \Vnoise{\nopix}
\end{equation}
where
$\Vabond{\nopix}=\left[\abond{1}{\nopix},\dots,\abond{\nbmat}{\nopix}\right]^T$
denotes the proportions of the $\nbmat$ materials in the $\nopix$th
pixel, $\MATmat=\left[\Vmat{1},\ldots,\Vmat{\nbmat}\right]$ is the
endmember matrix and $\Vnoise{\nopix}$ stands for an additive
residual term accounting for the measurement noise and modeling
error. Since the mixing coefficients
$\abond{1}{\nopix},\ldots,\abond{\nbmat}{\nopix}$ are expected to
represent the actual spatial distribution of the materials in the
$\nopix$th pixel, they are commonly subject to the following
positivity and sum-to-one (or additivity) constraints
\begin{equation}
  \label{eq:constraints_LMM}
  \left\{
  \begin{array}{ll}
    \abond{\nomat}{\nopix}\geq 0, & \forall \nomat,\ \forall \nopix  \\
    \sum_{\nomat=1}^{\nbmat} \abond{\nomat}{\nopix}=1, & \forall \nopix . \\
  \end{array}
  \right.
\end{equation}
This LMM has received a considerable attention in the image
processing and remote sensing literature since it represents an
acceptable first-order approximation of the physical processes
involved in most of the scenes of interest \cite{Keshava2002}.
Consequently, it has motivated a lot of research works that aim at
developing efficient endmember extraction algorithms (EEA), designed
to recover pure spectral signatures in the image, and inversion
techniques to estimate the abundance coefficients for a given
(estimated or \emph{a priori} known) set of endmembers.
Comprehensive overviews of these EEA and inversion methods can be
found in \cite{Keshava2002,Somers2011,Bioucas2012jstars}.
Specifically, two main approaches have been advocated to solve the
inversion step, that can be formulated as a constrained optimization
problem solved by fully constrained least square algorithms
\cite{Heinz2001,Theys2009,Heylen2011tgrs} or as a statistical
estimation problem solved within a Bayesian framework
\cite{Dobigeon2008,Eches2010ip,Eches2011icassp}.

However, for specific applications, LMM has demonstrated some
difficulties to accurately describe real mixtures
\cite{Dobigeon2014}. Notably, intimate mixtures of minerals are characterized by spatial scales
typically smaller than the path length followed by the photons, which violates one fundamental assumption for considering a linear model. Analyzing such mixtures, e.g., composed of minerals, requires to resort to complex physical models coming from the radiative transfer theory. Various approximating models have been proposed in the spectroscopic literature, such as the popular Hapke's model \cite{Hapke1981}. More recently, this model or related alternatives have been exploited in the hyperspectral literature to derive unmixing algorithms dedicated to remotely sensed images \cite{Guilfoyle2001,Nascimento2010}. Broadwater \emph{et al.} derived various kernel-based unmixing techniques that implicitly relied on the Hapke model \cite{Broadwater2009whispers,Broadwater2010whispers,Broadwater2011whispers}. In \cite{Close2012spie,Close2012spieb,Heylen2014}, the authors combined linear and intimate mixing processes in single models to improve flexibility.

Conversely, scenes acquired over vegetated
areas are also known to be subjected to more complex interactions that
can not be properly taken into account by a simple LMM
\cite{Borel1994,Ray1996,Zhang1998,Chen2006,Fan2009,Somers2009,Tits2012igarss,Somers2014}. Indeed, for these
specific scenarios addressed in this paper, differences in elevation between
the transparent $3$D vegetation canopies and the relatively flat
soil surfaces submit photons to multipath and scattering effects. Similar interaction effects have been also encountered when
analyzing urban scenes \cite{Huard2011whispers,Fontanilles2011,Meganem2014}.
Therefore, various attempts have been conducted to overcome the
intrinsic limitations of the LMM. A large family of nonlinear models
that have been proposed to analyze vegetated areas can be described
as
\begin{equation}
\label{eq:NLMM}
  \Vpix{\nopix} = \MATmat\Vabond{\nopix} +
  \boldsymbol{\mu}\left(\MATmat,\Vabond{\nopix},\Vnabond{\nopix}\right)
  + \Vnoise{\nopix}.
\end{equation}
In \eqref{eq:NLMM}, the observed pixel is composed of a linear
contribution similar to the LMM and an additive nonlinear term
$\boldsymbol{\mu}\left(\cdot\right)$ that may depend on the
endmember matrix $\MATmat$, the abundance coefficients in
$\Vabond{\nopix}$ and additional nonlinearity coefficients
$\Vnabond{\nopix}$ introduced to adjust the amount of nonlinearity
in the pixel. This class of models includes the bilinear models
\cite{Altmann2011whispers}, the quadratic-linear model
\cite{Meganem2014}, the post-nonlinear model \cite{Altmann2012ip}
and the bilinear-bilinear model \cite{Eches2014grsl} (the most
commonly used will be fully described in Section \ref{sec:models}).

However, to our knowledge, most of these models have been derived following physical or intuitive considerations, without any careful and thorough analysis of their
ability to properly describe real mixtures while performing spectral unmixing. In this article, we
propose to fill this gap by evaluating the relevance of various
nonlinear models when used for spectral unmixing of images acquired over vegetated areas. Specifically, requirements for ensuring the
quality of a model in this specific applicative context are threefold: \emph{(i)} the model should not depend on external parameters related to the studied scene (e.g., leaf index area, geometry or illumination incidence) since this prior knowledge is generally not available, \emph{(ii)} this model should be still sufficiently
flexible to fit the real observations in various external conditions, despite the ignorance of these unknown external parameters, \emph{(iii)} it should be able to account for the relative spatial
distribution of the materials in the pixel, with the prime objective
to estimate the abundance coefficients. In particular, mainly because of the two first requirements enounced above, advanced nonlinear models proposed in the remote sensing literature (e.g., \cite{Borel1994}) will not be considered in this study since they need a detailed prior knowledge regarding the analyzed scene.

To meet this challenge, we
take advantage of an interesting set of simulated and in-situ
collected hyperspectral data. First, we use a detailed virtual
orchard and forest model constructed in a physically based
ray-tracing environment using detailed sub-models for the
description of tree geometry, leaf and soil bidirectional
reflectance and diffuse illumination
\cite{Stuckens2009,VanderZande2010,Tits2012}. The model has been
thoroughly validated with field observations \cite{Stuckens2009},
and more recently we could provide, based on a comparison with
in-situ data, strong evidence that our ray tracing model
realistically describes the spectral scattering and thus
nonlinearity observed in vegetated areas \cite{Somers2014}. Second,
we use data from an in-situ experiment in a commercial citrus
orchard. This experiment comprised in-situ measured mixed pixel
reflectance spectra, pixel specific endmember spectra and subpixel
cover fraction distributions. This unique dataset of in-situ
measured mixed pixel reflectance spectra has previously been used to
study nonlinearity in fruit orchards \cite{Somers2009}.

The paper is organized as follows.  Section \ref{sec:models}
introduces the main nonlinear models that have been proposed in the
literature to describe mixtures encountered in vegetated areas. The
ray-tracer based simulated data and the in-situ measurements used to
validate these models are described in Section \ref{sec:data}. The
experiment results obtained by using the previously introduced
nonlinear models on the two sets of data are reported in Section
\ref{sec:results}. A comprehensive discussion on these results is
conducted in Section \ref{sec:discussion}. Section
\ref{sec:conclusion} concludes the paper.

\section{Nonlinear mixing models}\label{sec:models}
\subsection{Bilinear models}
\label{subsec:bilinear}
To take into account the scattering effects the photons are
subjected to before reaching the sensor, a wide class of nonlinear
models are derived by defining the nonlinear component
$\boldsymbol{\mu}\left(\MATmat,\Vabond{\nopix},\Vnabond{\nopix}\right)$
in \eqref{eq:NLMM} as a sum of bilinear terms
\cite{Altmann2011whispers}
\begin{equation*}
  \boldsymbol{\mu}\left(\MATmat,\Vabond{\nopix},\Vnabond{\nopix}\right)
  \triangleq \sum_{i=1}^{R-1}\sum_{j=i+1}^{R} \nabond{i,j}{\nopix} \Vmat{i}\odot \Vmat{j}
\end{equation*}
where the operator $\odot$ stands for a termwise product
\begin{equation}
\label{eq:term_BLMM}
  \Vmat{i}\odot \Vmat{j} \triangleq \left(
                                      \begin{array}{c}
                                        \mat{1}{i}\mat{1}{j} \\
                                        \vdots \\
                                        \mat{\nbband}{i}\mat{\nbband}{j} \\
                                      \end{array}
                                    \right).
\end{equation}
The set of nonlinearity coefficients
$\left\{\nabond{i,j}{\nopix}\right\}_{i,j}$ allows the amount of
nonlinearity in the $p$th pixel to be adjusted between each pair of
materials $\Vmat{i}$ and $\Vmat{j}$. Most of the various bilinear models of
the literature mainly differ by the definition of these coefficients
$\nabond{i,j}{\nopix}$ and the associated constraints they are
subject to. The most common models, that will be evaluated in
Section \ref{sec:results}, are recalled below.

In \cite{Somers2009} and \cite{Nascimento2009spie}, the authors
propose to include the nonlinearity coefficients
$\left\{\nabond{i,j}{\nopix}\right\}_{i,j}$ within the set of
constraints  \eqref{eq:constraints_LMM} defined by the LMM, leading
to
\begin{equation}
  \label{eq:FM}
  \Vpix{\nopix}^{(\textrm{NM})} \triangleq \sum_{\nomat=1}^{\nbmat} \abond{\nomat}{\nopix}
  \Vmat{\nomat} + \sum_{i=1}^{R-1}\sum_{j=i+1}^{R} \nabond{i,j}{\nopix} \Vmat{i}\odot \Vmat{j} + \Vnoise{\nopix}
\end{equation}
with
\begin{equation}
  \label{eq:constraints_NLMM}
  \left\{
   \begin{array}{l}
     \abond{\nomat}{\nopix}\geq 0, \quad \forall \nomat,\ \forall \nopix \\
     \nabond{i,j}{\nopix}\geq 0, \quad \forall \nomat,\ \forall i\neq j \\
     \sum_{r=1}^{\nbmat} \abond{\nomat}{\nopix}+
     \sum_{i=1}^{\nbmat-1}\sum_{j=i+1}^{\nbmat} \nabond{i,j}{\nopix}
     = 1, \ \ \forall \nopix.
   \end{array}
  \right.
\end{equation}

Note that this model, denoted NM for Nascimento's model in this
article, reduces to the LMM when $\nabond{i,j}{\nopix}= 0$, $\forall
i\neq j$. This is an interesting property since the LMM is known to
be an admissible first approximation of the actually involved
physical processes\footnote{It is widely admitted that the pixel spectrum measured by the sensor can be accurately described by the LMM when \emph{(i)} the photons are not subjected to multipath effects and \emph{(ii)} the materials are arranged side-by-side in the scene (as a checkerboard structure) \cite{Keshava2002}.}. However, in a more general case (i.e., $\nabond{i,j}{\nopix}\neq0$), the abundance
coefficients $\left\{\abond{r}{\nopix}\right\}_{r=1}^R$ are not
subject to the sum-to-one constraints defined in
\eqref{eq:constraints_LMM}.

In \cite{Fan2009}, Fan \emph{et al.} have defined  the nonlinearity
coefficients $\nabond{i,j}{\nopix}$ as the product of the
abundances,  $\nabond{i,j}{\nopix}\triangleq \abond{i}{\nopix}
\abond{j}{\nopix}$, under the LMM-based constraints in
\eqref{eq:constraints_LMM}, leading to the so-called Fan's Model
(FM)
\begin{equation}
  \label{eq:FM}
  \Vpix{\nopix}^{(\textrm{FM})} \triangleq \sum_{\nomat=1}^{\nbmat} \abond{\nomat}{\nopix}
  \Vmat{\nomat} + \sum_{i=1}^{R-1}\sum_{j=i+1}^{R} \abond{i}{\nopix} \abond{j}{\nopix} \Vmat{i}\odot \Vmat{j} +
  \Vnoise{\nopix}.
\end{equation}
The main motivation for relating the amount of nonlinear
interactions (governed by $\nabond{i,j}{\nopix}$) to the amount of
linear contribution (governed by $\abond{i}{\nopix}$ and
$\abond{j}{\nopix}$) is straightforward: the more a given material
is present in the pixel, the more nonlinear interactions may occur.
In particular, if a component $\Vmat{i}$ is absent in the $p$th
pixel, then $\abond{i}{\nopix} = 0$ and consequently
$\nabond{i,j}{\nopix}=0$, which means that there are no interactions
between the material $\Vmat{i}$ and any other materials $\Vmat{j}$
($j\neq i$). Note however that this bilinear model does not extend
the LMM.

To cope with this latter limitation, the generalized bilinear model
(GBM) \cite{Halimi2011} weights the products of abundances
$\abond{i}{\nopix} \abond{j}{\nopix}$ by additional free parameters
$\gamma_{i,j,p}\in (0,1)$ that tune the amount of nonlinear
interactions, leading to $\nabond{i,j}{\nopix} \triangleq
\gamma_{i,j,p}\abond{i}{\nopix} \abond{j}{\nopix} $ and
\begin{equation}
  \label{eq:GBM}
  \Vpix{\nopix}^{(\textrm{GBM})} \triangleq \sum_{\nomat=1}^{\nbmat} \abond{\nomat}{\nopix}
  \Vmat{\nomat} + \sum_{i=1}^{R-1}\sum_{j=i+1}^{R} \gamma_{i,j,p}\abond{i}{\nopix} \abond{j}{\nopix} \Vmat{i}\odot \Vmat{j} +
  \Vnoise{\nopix}.
\end{equation}
The GBM has the nice properties of i) generalizing the LMM by
enforcing $\gamma_{i,j,p}=0$ ($\forall i,j$), similarly to NM but
contrary to FM and ii) having the amount of nonlinear interactions
to be proportional to the material abundances, similarly to FM but
contrary to NM.

\subsection{Post-nonlinear mixing model}
Inspired by pioneered works in blind source separation
\cite{Taleb1999}, Altmann \emph{et al.} have introduced in
\cite{Altmann2012ip} a nonlinear model that relies on a $2$nd-order
polynomial expansion of the nonlinearity,
\begin{equation}\label{eq:term_PPNM}
\begin{split}
  \boldsymbol{\mu}\left(\MATmat,\Vabond{\nopix},\Vnabond{\nopix}\right)
&\triangleq b_p \left(\MATmat\Vabond{p}\right) \odot
\left(\MATmat\Vabond{p}\right)\\
&=  b_p \sum_{i=1}^R\sum_{j=1}^R \abond{i}{\nopix} \abond{j}{\nopix} \Vmat{i}\odot \Vmat{j}
\end{split}
\end{equation}
leading to the following polynomial post-nonlinear mixing model
(PPNM)
\begin{equation}
  \Vpix{\nopix}^{(\textrm{PPNM})} = \MATmat\Vabond{p} + b_p \sum_{i=1}^R\sum_{j=1}^R \abond{i}{\nopix} \abond{j}{\nopix} \Vmat{i}\odot \Vmat{j}+ \Vnoise{p}.
\end{equation}
The PPNM has demonstrated a noticeable flexibility to model various
nonlinearities not only for unmixing purposes \cite{Altmann2012ip}
but also to detect nonlinear mixtures in the observed image
\cite{Altmann2013ip}. This model has also the great advantage of
having the amount of nonlinearity to be governed by a unique
parameter $b_p$ in each pixel, contrary to NM or GBM.
Eq. \eqref{eq:term_PPNM} also shows PPNM includes bilinear
terms $\Vmat{i}\odot \Vmat{j}$ ($j \neq i$) similar to those
involved in the NM, FM and GBM, and also quadratic terms
$\Vmat{i}\odot \Vmat{i}$, which may account for interactions between
similar materials.

\subsection{Unmixing algorithms}
\label{subsec:algorithm} To evaluate the accuracy of the mixing
models of interest, the pixels of the in-situ and simulated data are
unmixed with respect to each model. When analyzing the pixels with
the LMM, the nonlinear contribution
$\boldsymbol{\mu}\left(\MATmat,\Vabond{\nopix},\Vnabond{\nopix}\right)$
is set to zero. Based on the prior knowledge of the endmember
signatures $\MATmat$, the abundance vector $\Vabond{\nopix}$
associated with each pixel $\Vpix{\nopix}$ is estimating by solving
the constrained minimization problem
\begin{equation}
  \label{eq:criterion_LMM}
  \hatVabond{\nopix} =
  \operatornamewithlimits{argmin}_{\Vabond{\nopix}}
  \left\|\Vpix{\nopix} - \MATmat\Vabond{\nopix}\right\|_2^2 \quad
  \text{s.t.} \quad \eqref{eq:constraints_LMM}.
\end{equation}

In this work, to solve this problem, the fully constrained least
square (FCLS) algorithm \cite{Heinz2001} is used.

Moreover, when analyzing the pixels with nonlinear mixing models,
the abundance vector $\Vabond{\nopix}$ and the nonlinearity
parameter vector $\Vnabond{\nopix}$ associated with each pixel
$\Vpix{\nopix}$ are estimated by solving the following constrained
optimization problem
\begin{equation}
  \label{eq:criterion_NLMM}
  \left(\hatVabond{\nopix}, \hatVnabond{\nopix}\right) =
  \operatornamewithlimits{argmin}_{\Vabond{\nopix},\Vnabond{\nopix}}
  \left\|\Vpix{\nopix} - \MATmat\Vabond{\nopix} -
  \boldsymbol{\mu}\left(\MATmat,\Vabond{\nopix},\Vnabond{\nopix}\right)\right\|_2^2.
\end{equation}
Depending on the considered model, the set of constraints imposed to
the abundance vector $\Vabond{\nopix}$ and the possible nonlinear
coefficient vector $\Vnabond{\nopix}$ may differ. For the FM, GBM
and PPNM, the abundance vector $\Vabond{\nopix}$ should satisfy the
LMM-based constraints \eqref{eq:constraints_LMM}, while for the NM,
this constraint is applied to the joint vector
$\left[\Vabond{\nopix},\Vnabond{\nopix}\right]$. Similarly, the
nonlinear coefficient vector $\Vnabond{\nopix}$ for the GBM and PPNM
should satisfy constraints that depend on the considered model and
the nonlinearity component
$\boldsymbol{\mu}\left(\MATmat,\Vabond{\nopix},\Vnabond{\nopix}\right)$
in \eqref{eq:term_BLMM} or \eqref{eq:term_PPNM} depends also on the
considered nonlinear model.

For the experimental results reported in Section~\ref{sec:results},
the FCLS algorithm is used to solve the NM-based unmixing problem
since NM can be interpreted as a linear mixture of an extended set
of endmembers, as shown in \cite{Nascimento2009spie}. The FM
parameters are estimated with the algorithm detailed in
\cite{Fan2009}, based on a first-order Taylor series expansion of
the nonlinearity
$\boldsymbol{\mu}\left(\MATmat,\Vabond{\nopix},\Vnabond{\nopix}\right)$.
Finally, the gradient descent and the subgradient descend algorithms
developed in \cite{Halimi2011igarss} and \cite{Altmann2012ip} are
used to solve the GBM- and PPNM-based unmixing problems,
respectively. Interested readers are invited to refer to these works
for detailed information regarding the optimization schemes.

\section{Data description}\label{sec:data}

The mixing models and corresponding unmixing algorithms detailed in
the previous sections are compared using simulated and real
hyperspectral images. It is worth noting that, for both kinds of
datasets, actual pure component spectral signatures (i.e., endmember
spectra) and quantitative spatial distributions of these components
(i.e., abundances) are available as ground truth in each pixel of
the considered images. These datasets\footnote{They will be
available online at
\url{http://www.biw.kuleuven.be/m3-biores/geomatics/data/}.} are
described in this section.


\subsection{Simulated dataset}

Two types of synthetic hyperspectral image data were generated from
a ray tracing experiment. First, synthetic but realistic fully calibrated virtual scenes, namely citrus orchards and a forest, have
been designed using methods developed in \cite{Stuckens2009} and
\cite{VanderZande2008}, respectively, which will be explained in more detail in the following paragraphs. Then, corresponding
hyperspectral images have been simulated using an extended version
of the physically based ray tracer (PBRT) \cite{Pharr2004}. In PBRT,
a scene is defined using submodels to describe the various
components of the scene: illumination sources, sensor platform,
material optical properties, integrator and geometry descriptions.
For the different generated images, the illumination has been
modelled to closely agree with the average circadian illumination
from April until September, corresponding to a midlatitude northern
hemisphere growing season. The illumination has been composed of a
combination of direct and diffuse light calculated from $350$ to
$2500$nm with a $10$nm interval. The citrus trees and weeds of the
orchard scenes (see paragraph \ref{sec:orchard}) and the trees of
the forest scene (see paragraph \ref{sec:forest}) have been
constructed as triangular meshes by implementing the algorithm
introduced in \cite{Weber1995}. Their material properties have been
described by a bidirectional scattering distribution function (BSDF)
model \cite{Stuckens2009}.

\subsubsection{Orchard scenes}\label{sec:orchard}

\begin{figure}[h!]
    \centering
        \includegraphics[width=0.75\columnwidth]{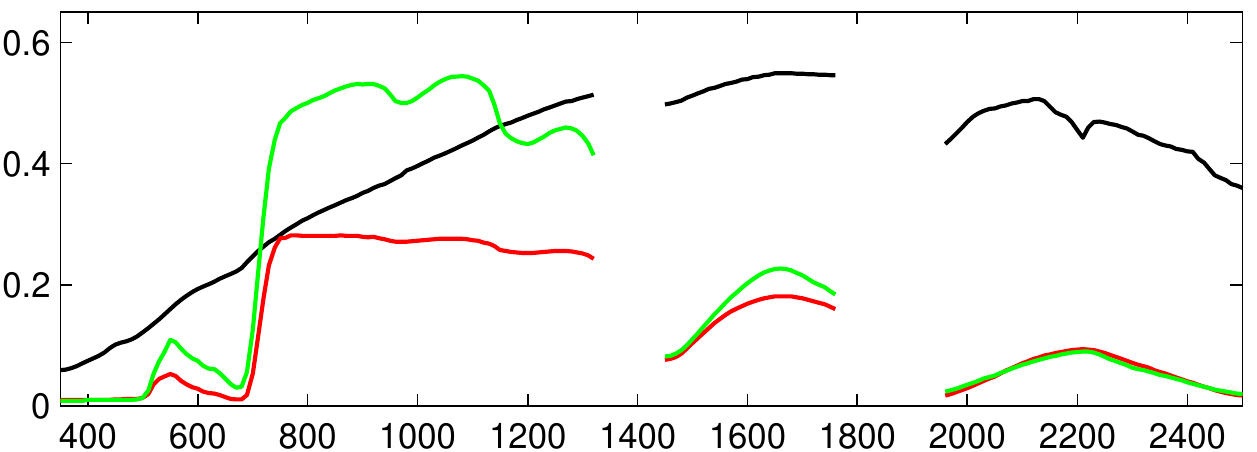}
        \caption{Two- and three-endmember orchard synthetic dataset. Endmember spectra: soil (black), weed (red) and tree (green).}
        \label{fig:orchard_endmembers}
\end{figure}

The fully calibrated virtual citrus orchard developed in
\cite{Stuckens2009} has been used to create two different orchard
scenes: (\textit{i}) an orchard consisting of citrus trees and a
soil background, leading to two-endmember mixtures and (\textit{ii})
an orchard consisting of citrus trees, a soil background and weed
patches, leading to three-endmember mixtures. Each orchard scene
consists of $20\times20$ pixels, with a pixel size of
$2\textrm{m}\times2\textrm{m}$. The exact per-pixel
abundances are known for the three components, as well as the reference spectral signatures. More precisely,
for the soil endmember, the pure spectral signature consisted of the fully sunlit soil uncontaminated by
the surrounding trees. For the tree endmember, the soil background of the orchard was replaced by a perfectly
absorbing background, to minimize the influence of the background on the tree signature. A $5$cm
resolution image of $4$m by $4$m was rendered above a canopy with one row. Only the pixels containing a
tree fraction greater than $0.95$ were retained and averaged to provide the pure tree signature. As such,
the tree spectral signatures are an integration of all components of a tree, including sunlit and shaded
leafs, branches and stems. For the weed endmember spectral signature, a similar approach to the tree
endmember was used, replacing the soil background with a perfectly absorbing background, and removing
all trees from the orchard. A $4$m by $4$m image with $1$cm resolution was rendered over a weed patch,
selecting only those pixels with a weed fraction greater than $0.95$. Finally, these pixels were averaged
to provide the spectral signature of the weeds. The resulting endmember spectra
are depicted in Fig.~\ref{fig:orchard_endmembers}.

\begin{figure}[h!]
    \centering
    \begin{subfigure}{0.49\columnwidth}
        \centering
        \includegraphics[width=\columnwidth]{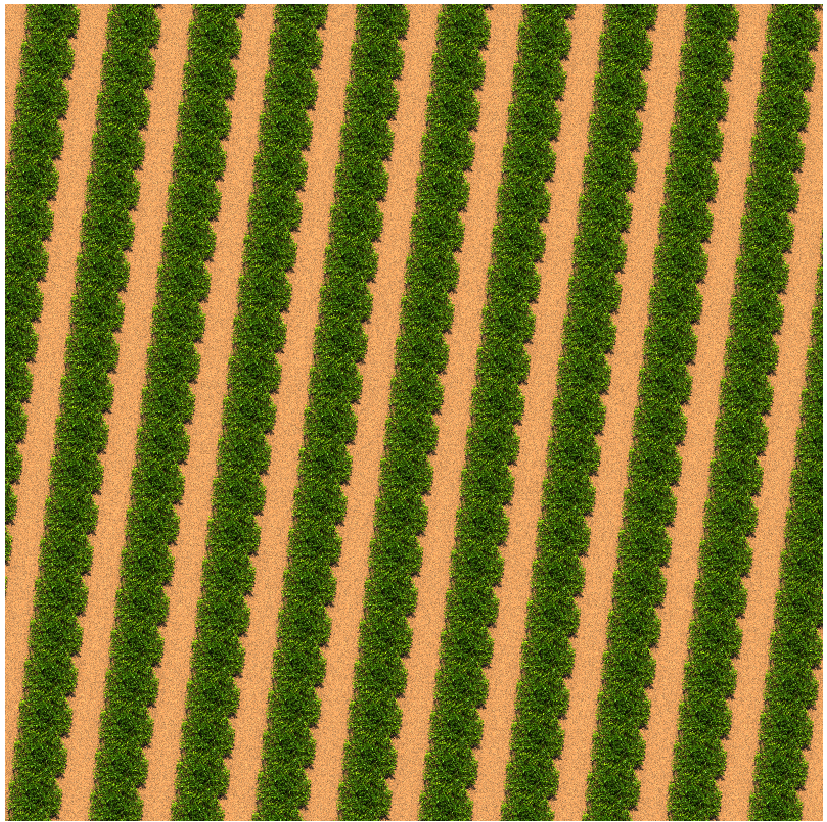}
        \caption{}
    \end{subfigure}
\begin{subfigure}{0.49\columnwidth}
    \centering
    \includegraphics[width=\columnwidth]{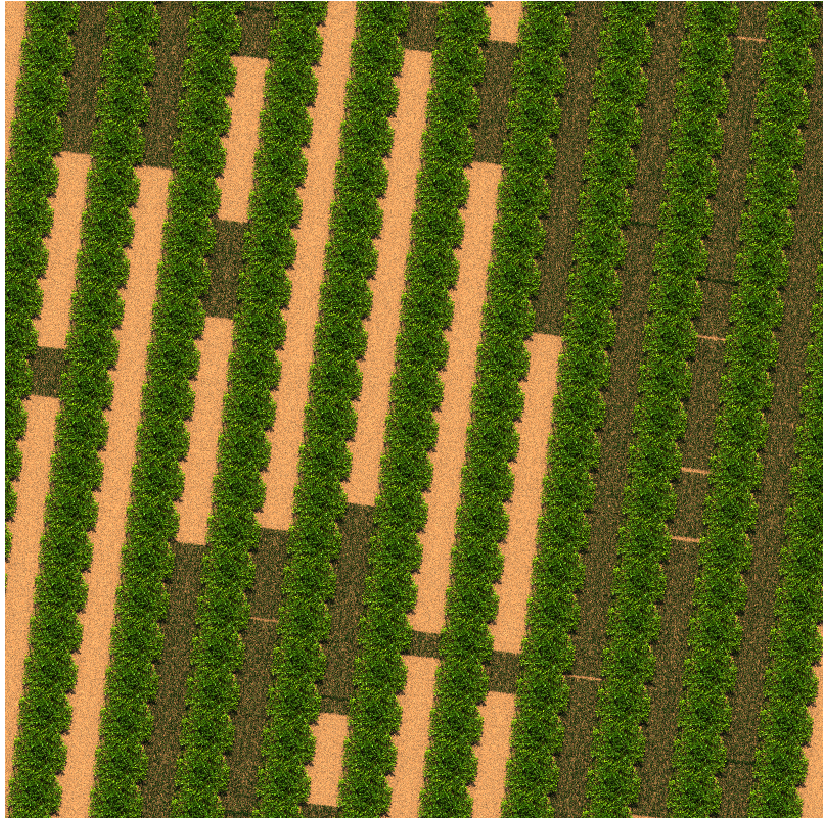}
    \caption{}
\end{subfigure}
\caption{High resolution images of the two orchards with (a) two
endmembers, i.e., tree and soil, and (b) three endmembers, i.e.,
tree, soil and weeds. } \label{fig:orchardExample}
\end{figure}

The orchards have been constructed with a row spacing of $4.5$m,
tree spacing of $2$m, row azimuth of $7.3^{\circ}$  and an average
tree height of $3$m. This composition is consistent with the
reference orchard, located in Wellington, South Africa
($33.58^{\circ}$S, $18.93^{\circ}$E), used to calibrate the virtual
orchard \cite{Stuckens2009}. Spectral input data for citrus leaves
and stems, soils and weeds have been measured using a full-range
($350$-$2500$nm) analytic spectral devices (ASD) Fieldspec JR
spectroradiometer with a $25^{\circ}$ foreoptic. The weed spectrum
has been chosen as of the Lolium sp. species. A Haplic Arenosol
\cite{FAO1998} typical for commercial citrus orchards in the Western
Cape Province in South Africa has been used in the simulations
\cite{Somers2010a}. An example of a high resolution image of
$20\textrm{m} \times20\textrm{m}$ of the two-endmember orchard is
depicted in Fig~\ref{fig:orchardExample} (a), while the
three-endmember orchard is shown in Fig~\ref{fig:orchardExample}
(b). For a detailed description of the design, modalities and
application of the virtual orchard, the reader is invited  to
consult \cite{Stuckens2009}.

\subsubsection{Forest scene}\label{sec:forest}

\begin{figure}[h!]
  \centering
  \includegraphics[width=0.75\columnwidth]{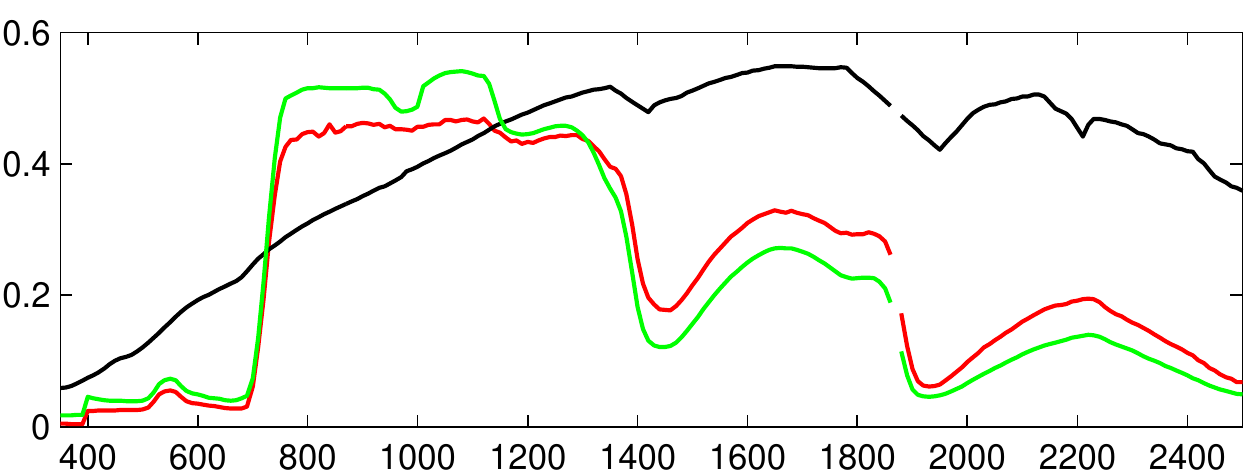}
   \caption{Forest synthetic dataset. Example of the generated endmember spectra: soil (black), beech (red) and pop (green).}
  \label{fig:forest_endmembers}
\end{figure}

The virtual forest consisted of a soil background planted with trees selected from the
species-specific tree pools. More precisely, to simulate the forest scene, $3$D tree geometry descriptions were
available for beech (\textit{Fagus sylvatica} L.) and poplar (\textit{Populus nigra}
L. var. \textit{"italica"} Muench). Each tree was characterized by a
specific structure based on its age (i.e., $20$ years old). All
leaves were assigned a species-specific reflectance and
transmittance spectrum extracted from the leaf optical properties
experiment (LOPEX) dataset \cite{Hosgood1995}. Examples of the soil, beech and pop endmember signatures are depicted in
Fig~\ref{fig:forest_endmembers}.

\begin{figure}[h!]
  \centering
  \includegraphics[width=0.75\columnwidth]{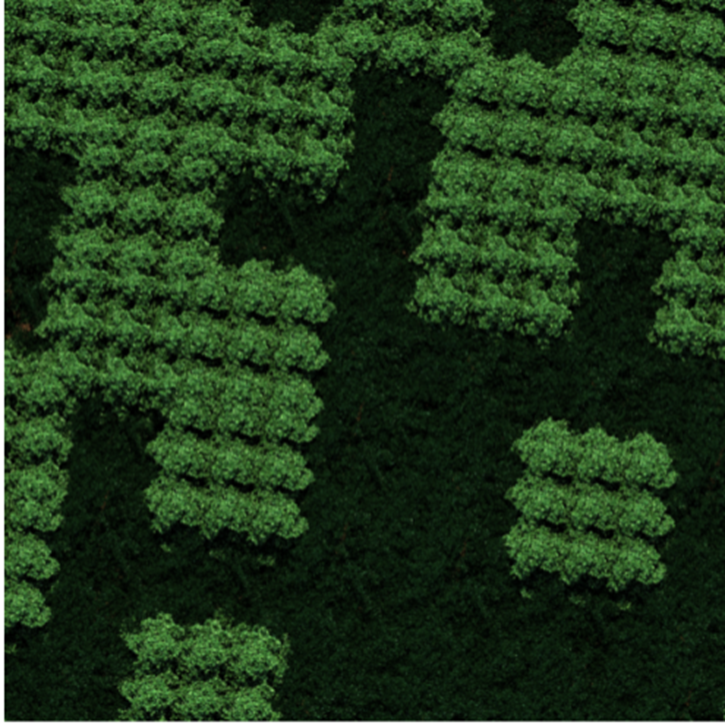}
  \caption{High resolution detail of a $30$m pixel of the forest with 60$\%$ beech trees and 40$\%$ poplars.}
  \label{fig:forestExample}
\end{figure}

To achieve a nearly
$100\%$ canopy cover, the average tree spacing has been set to $5$m
for the beech trees and $1$m for the poplars. A series of six forest
scenes has been rendered providing a gradual transition from a
forest scene completely dominated by one species to a scene
dominated by the other species. More precisely, $20\%$ of the beech
trees have been randomly replaced by poplar trees in the subsequent
scene. Each forest scene consisted of $15\times15$ pixels, with a
pixel size of $30 \textrm{m} \times 30 \textrm{m}$. In
Fig~\ref{fig:forestExample}, a detail is shown of a $30$m pixel, for
the forest consisting of 60$\%$ beech trees and 40$\%$ poplars. Note
here that the spatial resolution of the forest scene is
significantly larger than the resolution of the orchard scene
detailed in paragraph \ref{sec:orchard}. These choices allow
different plant production systems to be covered, with various
species combinations, sets of endmembers and spatial resolution
scales.

\subsection{In-situ measurement} \label{sec:in situ}

\begin{figure}[h!]
  \centering
  \includegraphics[width=0.75\columnwidth]{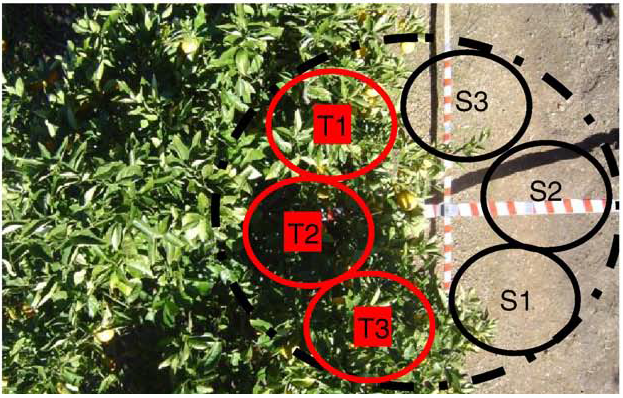}
   \caption{Experimental set-up to determine plot-specific soil and tree endmember signatures for each plot (from \cite{Somers2009}). The areas T1, T2 and T3 (S1, S2, and S3, resp.)
   identify the sub-plots selected for the measurements of pure tree (soil, resp.) spectra. These measurements are averaged to provide the plot-specific tree (soil, resp.)
   endmember signature.}
  \label{fig:setup}
\end{figure}

\begin{figure}[h!]
  \centering
  \includegraphics[width=0.75\columnwidth]{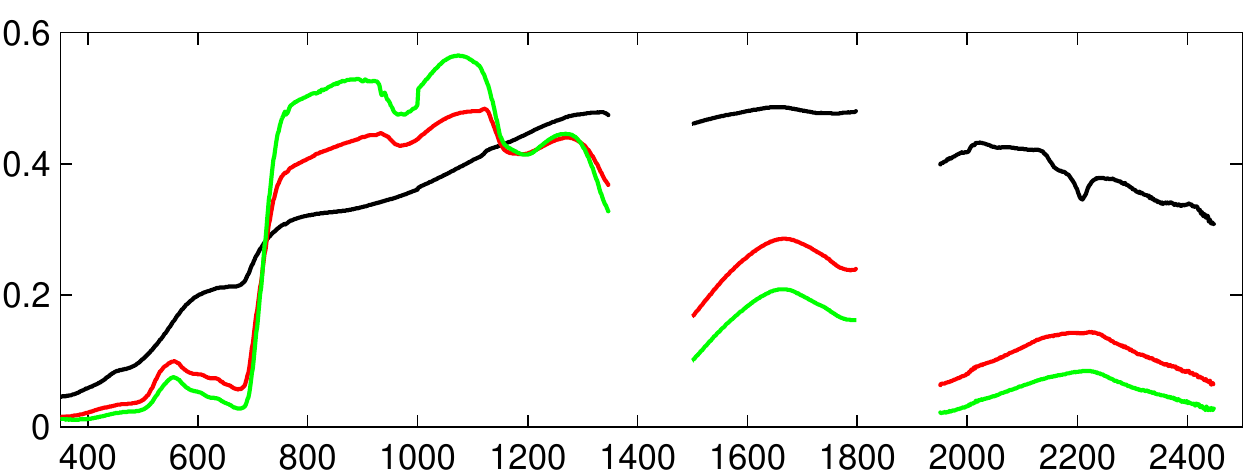}
   \caption{Two- and three-endmember in-situ measurements. Example of the measured endmember spectra: soil (black), weed (red) and tree (green).}
  \label{fig:insitu_endmembers}
\end{figure}

In addition, an experiment was conducted in the same orchard used for the calibration
of the virtual orchard described in paragraph \ref{sec:orchard}. Significant weed cover,
dominantly \emph{Lolium sp. L.} ($\approx30\%$ of the inter-row spacing, concentrated in dense patches) was present. Throughout the orchard, in-situ measured reflectance spectra of $60$
mixed ground plots were collected, i.e., $25$ mixtures of tree and
soil, $25$ mixtures of tree and weed, and $25$ mixtures of tree,
soil and weed. Reflectance measurements were performed in August
using a spectroradiometer with a $25^{\circ}$ fore-optic, covering
the $350-2500$nm spectral domain (Analytic Spectral Devices,
Boulder, USA). The measurements were taken from nadir at a height of
$4$m. For each measured mixed pixel, the plot-specific pure
endmember spectra and ground cover fraction distributions were
determined. Specifically, to mitigate the impact of nonlinear mixing from endmember variability, plot-specific endmembers were acquired by measuring
a number of pure spectra in each plot, as
illustrated in Fig. \ref{fig:setup}. One set of soil, weed and tree endmember spectra is depicted in Fig.~\ref{fig:insitu_endmembers}.

Information on the ground cover
composition of each of the measured mixed pixels was extracted from
digital photographs (SONY DSC-P8/3.2 megapixel cyber shot camera,
positioned in nadir). A more detailed description on the
experimental setup, depicted in Fig.~\ref{fig:setup}, can be found
in \cite{Somers2009}.

\section{Experimental results}

The relevance of the mixing models under test, namely LMM, FM, NM,
GBM and PPNM, and associated unmixing algorithms, is evaluated with
respect to i) their ability of accurately describing the physical
processes yielding the considered mixtures and ii) their ability of
providing meaningful estimations of the abundance coefficients, to
properly account for the spatial distribution of the materials over
each observed pixel. More precisely, let $\hatVabond{\nopix}$ and
$\hatVnabond{\nopix}$ denote the abundance and nonlinearity
coefficient vectors estimated by the algorithms introduced in
paragraph \ref{subsec:algorithm}. First, the average square
reconstruction error (RE) is measured as
\begin{equation}
\label{eq:RE}
 \mathrm{RE} = \frac{1}{\nbband\nbpix} \sum_{\nopix=1}^{\nbpix} \left\|\Vpix{\nopix}-\hatVpix{\nopix}\right\|^2
\end{equation}
where $\left\|\cdot\right\|$ stands for the usual Euclidean norm
($\left\|\bfx\right\| = \sqrt{\bfx^T\bfx}$). In the right-hand side
of \eqref{eq:RE}, $\Vpix{\nopix}$ ($\nopix=1,\ldots,\nbpix$) are the
observed pixels whereas $\hatVpix{\nopix}$ are the corresponding
estimates given by
\begin{equation}
\hatVpix{\nopix} = \MATmat\hatVabond{\nopix} + \boldsymbol{\mu}\left(\MATmat,\hatVabond{\nopix},\hatVnabond{\nopix}\right)
\end{equation}
where $\boldsymbol{\mu}\left(\cdot\right)$ is equal to $0$ for the
LMM or stands for the additional nonlinear contribution for the
nonlinear models (see Section \ref{sec:models}).

Since the actual endmember spectra and abundance coefficients (that
satisfy the constraints in \eqref{eq:constraints_LMM}) are perfectly
known for each pixel of the considered scenes, these REs can also be
computed from pixels reconstructed following the LMM and FM with the
actual values of the abundances. These two ``oracle'' models are
denoted o-LMM and o-FM in what follows. In particular, the RE
associated with the o-LMM provides interesting information regarding
the actual level of nonlinearities in the considered pixels. Note
also that such oracle performance can not be computed for the other
nonlinear models, since NM is based on a different abundance
definition (e.g., they do not follow the constraints
\eqref{eq:constraints_LMM}) and GBM and PPNM require the prior
knowledge of additional (unknown) parameters.

Moreover, to visualize the reconstruction error as a function of the wavelength,
a signed error, defined as the mean reconstruction difference in the
$\ell$th band, is also computed as
\begin{equation}
 \mathrm{RD}_{\ell} = \frac{1}{\nbpix} \sum_{\nopix=1}^{\nbpix}
 \left(\pix{\ell}{\nopix}-\hatpix{\ell}{\nopix}\right).
\end{equation}
Finally, to measure the accuracy of the abundance estimation, the
mean square errors (MSE) between the actual abundance vectors
$\Vabond{\nopix}$ and the corresponding estimated
$\hatVabond{\nopix}$ ($\nopix=1,\dots,\nbpix$) are computed as
follows
\begin{equation}
\mathrm{MSE} = \frac{1}{\nbmat\nbpix} \sum_{\nopix=1}^{\nbpix} \left\|\Vabond{\nopix}-\hatVabond{\nopix}\right\|^2.
\end{equation}

\label{sec:results}
\subsection{Simulated dataset}
\subsubsection{Virtual orchard}
\label{subsubsec:results_orchard}
The unmixing results for the simulated orchard scenes are shown in
Table \ref{tab:synth_orchard_RMSE} in terms of MSE and RE. From
these results, for both two- and three-endmembers, one can conclude
that NM and LMM perform similarly in term of RE, while PPNM and FM
provide the best results and, in particular, significantly better
than LMM. It is interesting  to note that, for the $2$-endmember mixtures, GBM does not provide smaller RE than LMM, as expected. Indeed, as highlighted in paragraph \ref{subsec:bilinear}, GBM reduces to LMM if $\gamma_{i,j,p}=0,\ \forall i,j$, which is supposed to confer to GBM more flexibility than LMM. This might indicate that the unmixing algorithm associated with GBM has not properly converged for this dataset. This point is discussed in more details in Section \ref{sec:discussion}. Regarding the abundance MSE, NM and LMM provide similar
errors for two-endmember mixtures and all nonlinear models perform
better than LMM for three-endmember mixtures.

\begin{table}[h!]
\renewcommand{\arraystretch}{1.15}
\begin{center}
\begin{tabular}{|c|c||c|c|}
\hline
\multicolumn{2}{|c||}{  }                           & $2$ endm.             & $3$ endm. \\
\hline
\multirow{7}*{\rotatebox{90}{RE}}           & LMM           & $7.70     $            &  $5.81    $          \\ 
                                            & o-LMM            & $15.0     $            &  $10.40    $            \\ 
                                            & FM            & $\best{1.24}     $            &  $\second{0.91}    $            \\ 
                                            & o-FM            & $10.20     $            &  $7.66    $            \\ 
                                            & NM            & $7.70     $            &  $5.81    $          \\ 
                                            & GBM           & $10.13     $            &  $0.94   $           \\ 
                                            & PPNM          & $\second{1.28}     $            &  $\best{0.91}    $              \\ 
\hline
\multirow{5}*{\rotatebox{90}{MSE}}         & LMM           & $\second{0.96}     $            &  $3.17    $            \\ 
                                            & FM            & ${1.13}     $            &  $\best{2.27}    $            \\ 
                                            & NM            & $\best{0.92}     $            &  $\second{2.44}    $            \\ 
                                            & GBM           & $1.47     $            &  ${2.45}    $         \\ 
                                            & PPNM          & ${1.22}     $            &  ${2.62}    $        \\ 
\hline
\end{tabular}
\caption{Two- and three-endmember orchard synthetic dataset.
Abundance MSE ($\times 10^{-2}$) and RE ($\times 10^{-4}$) for
various linear/nonlinear mixing
models.\label{tab:synth_orchard_RMSE}}
\end{center}
\end{table}

\begin{figure}[h!]
  \centering
  \includegraphics[width=0.75\columnwidth]{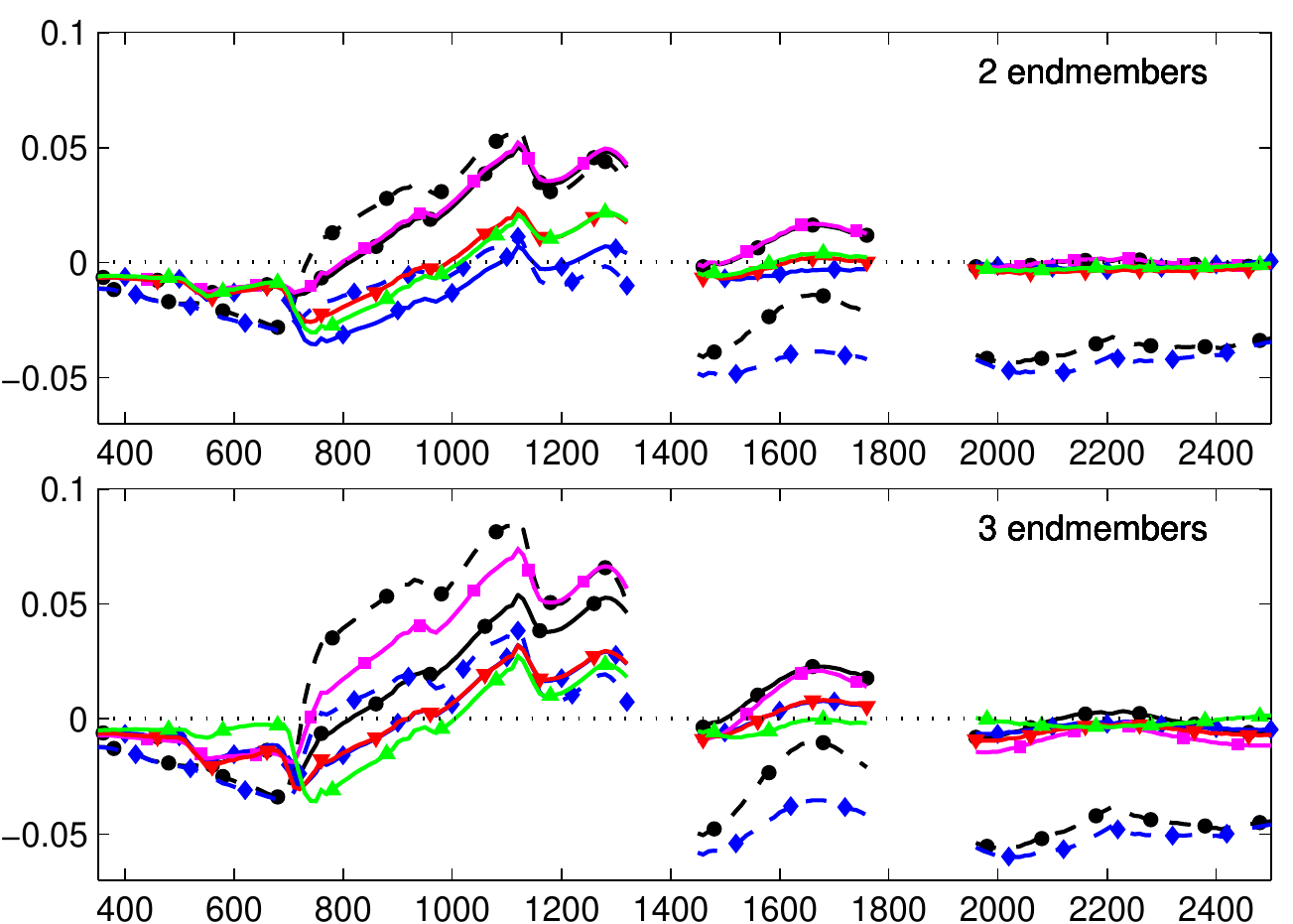}
  \caption{Two- and three-endmember orchard synthetic dataset. Reconstruction difference $\textrm{RD}_{\ell}$ as a function of wavelength for various linear/nonlinear mixing
models: LMM (black), oracle-LMM (black, dashed line), FM (blue),
oracle-FM (blue, dashed line), NM (magenta), GBM (red) and PPNM
(green).}
  \label{fig:synth_orchard_wavelength}
\end{figure}

In Fig. \ref{fig:synth_orchard_wavelength}, the RDs are depicted as
functions of wavelength, for the different linear and nonlinear
mixing models. From this figure, it appears that the nonlinearities
occurring in spectral bands ranging from $1400$nm to $2500$nm are of
high intensity (see the plot associated with the oracle-LMM, in
black dashed line) but are rather well described by the various
nonlinear models.

\subsubsection{Virtual forest}

For the simulated forest scenes, the unmixing results are reported
in Table \ref{tab:synth_forest_RMSE}. These results are computed for
four scene compositions, with increasing proportions from $20\%$ to
$80\%$ of beech trees with respect to poplars (see paragraph
\ref{sec:forest}). The first three images provided a sequence of
images with increasing nonlinearity, as shown by the RE obtained
with the oracle-LMM, ranging from $2.11$ to $5.36$ ($\times
10^{-4}$). The fourth image, composed of $80\%$ of poplars and
$20\%$ of beech trees, seems to be subject to nonlinearities of
lower intensity, since the oracle-LMM RE is $3.15 \times 10^{-4}$.

As with the previous dataset, NM together with PPNM provides the
best model fit for all images, i.e. with lowest RE, and the best
abundance estimates in terms of MSE. The abundance estimation
performance of the different models is also decreasing with
increasing nonlinear mixing effects in the images, even though the
RE remained almost constant for NM and PPNM. FM performed poorly and
LMM and GBM lead to similar results.

\begin{table}[h!]
\renewcommand{\arraystretch}{1.15}
\begin{center}
\begin{tabular}{|c|c||c|c|c|c|}
\hline
\multicolumn{2}{|c||}{$\% $beech}                           & $0.2$             & $0.4$         & $0.6$         & $0.8$         \\ 
\multicolumn{2}{|c||}{$\% $pop  }                           & $0.8$             & $0.6$         & $0.4$         & $0.2$         \\ 
\hline
\multirow{7}*{\rotatebox{90}{RE}}           & LMM           &	$	0.92	$            &  $	1.78	$            &  $	1.84	$            &  $	0.88	$            	\\
                                            & o-LMM         &	$	2.11	$            &  $	4.37	$            &  $	5.36	$            &  $	3.15	$            	\\
                                            & FM            &	$	1.37	$            &  $	3.33	$            &  $	4.56	$            &  $	3.39	$            	\\
                                            & o-FM          &	$	7.01	$            &  $	20.54	$            &  $	33.03	$            &  $	24.56	$            	\\
                                            & NM            &	$	\second{0.23}	$            &  $	\best{0.11}	$            &  $	\best{0.10}	$            &  $	\best{0.12}	$            	\\
                                            & GBM           &	$	0.92	$            &  $	1.78	$            &  $	1.84	$            &  $	0.88	$            	\\
                                            & PPNM          &	$	\best{0.13}	$            &  $	\second{0.15}	$            &  $	\second{0.14}	$            &  $	\second{0.12}	$            	 \\
\hline										
\multirow{5}*{\rotatebox{90}{MSE}}         & LMM           & 	$	0.73	$            &  $	2.44	$            &  $	4.98	$            &  $	3.18	$            	\\
                                            & FM            & 	$	1.43	$            &  $	5.16	$            &  $	11.65	$            &  $	14.95	$            	\\
                                            & NM            & 	$	\best{0.25}	$            &  $	\best{0.58}	$            &  $	\best{0.66}	$            &  $	\second{0.45}	$            	\\
                                            & GBM           & 	$	0.72	$            &  $	2.45	$            &  $	5.01	$            &  $	3.22	$            	\\
                                            & PPNM          & 	$	\second{0.40}	$            &  $	\second{0.80}	$            &  $	\second{0.93}	$            &  $	\best{0.62}	$            	 \\

\hline
\end{tabular}
\caption{Three-endmember forest synthetic dataset. Abundance MSE
($\times 10^{-2}$) and RE ($\times 10^{-4}$) for various
linear/nonlinear mixing models. \label{tab:synth_forest_RMSE}}
\end{center}
\end{table}

\begin{figure}[h!]
  \centering
  \includegraphics[width=0.75\columnwidth]{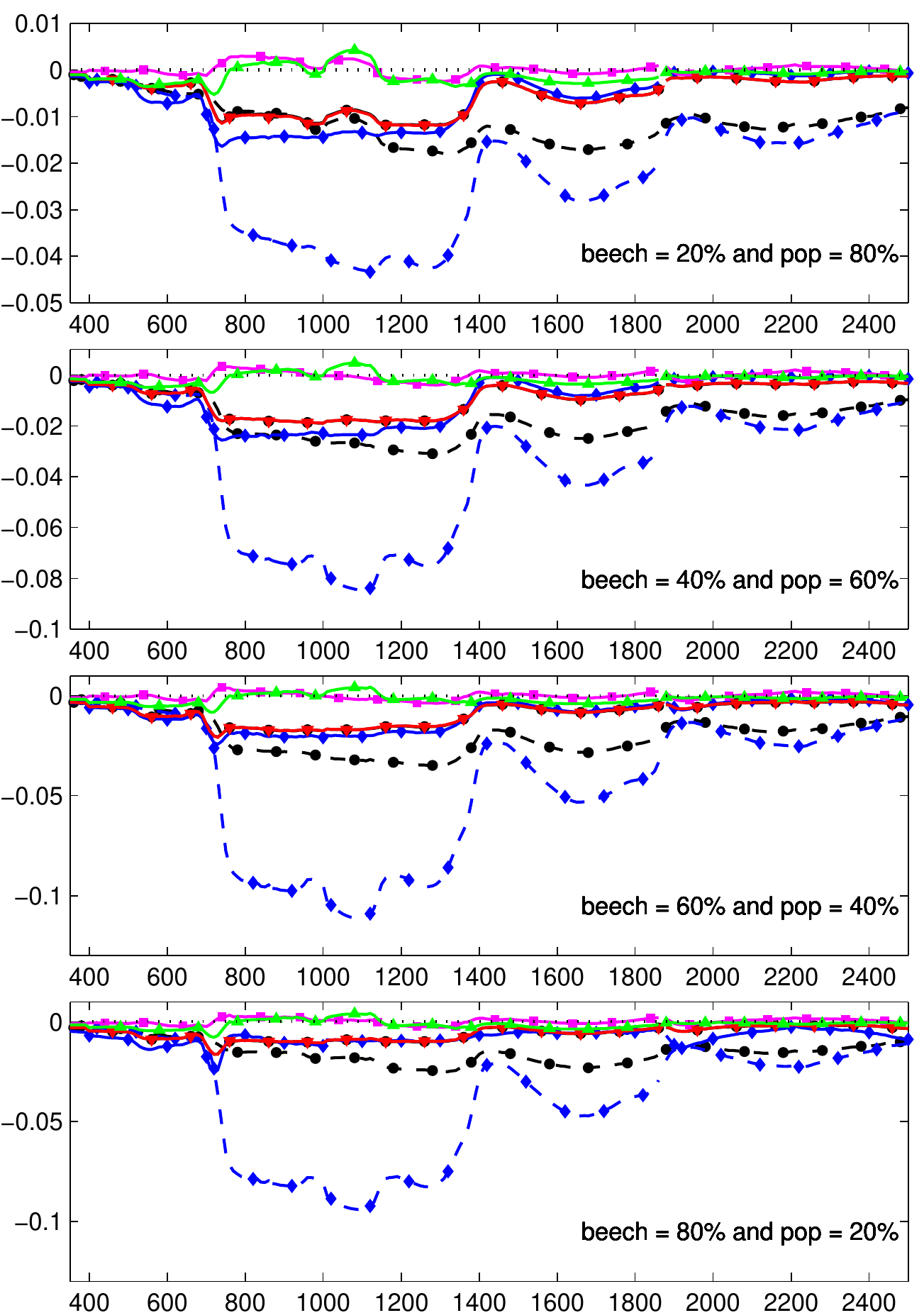}
  \caption{Three-endmember forest synthetic dataset. Reconstruction difference $\textrm{RD}_{\ell}$ as a function of wavelength for various linear/nonlinear mixing
models: LMM (black), oracle-LMM (black, dashed line), FM (blue),
oracle-FM (blue, dashed line), NM (magenta), GBM (red) and PPNM
(green).}
  \label{fig:synth_forest_wavelength}
\end{figure}

Fig. \ref{fig:synth_forest_wavelength} shows the RDs as functions of
wavelength. From the RD associated with the oracle-LMM, it clearly
appears that the nonlinearity effects mostly occur in the spectral
range $700\textrm{nm}-1400\textrm{nm}$, especially for the $20-80\%$
and $80-20\%$ scenes. All nonlinear mixing models provide good model
fits, except the FM, as already shown by the REs reported in Table
\ref{tab:synth_forest_RMSE}.

\subsection{In-situ measurements}
Three types of in-situ measured mixed pixels were available to test
the different mixing models, i.e., tree-weed, tree-soil and
tree-soil-weed mixtures (see paragraph \ref{sec:in situ}). In Table
\ref{tab:RMSE_real}, the reconstruction error of the mixed signal
and the accuracy of the estimated abundances are depicted. From the
RE associated with the oracle-LMM, it appears that most
nonlinearities occur in the tree-soil mixtures. Once again, PPNM is
the mixing model that reconstructs the mixed signatures the best,
while FM performed worse than the LMM. For the abundance accuracy,
MSE results are less homogeneous than those obtained with the
various simulated datasets. Depending on the type of the mixture,
GBM or PPNM are the best unmixing model, while FM gives the lowest
abundance estimation accuracies.

\begin{table}[h!]
\renewcommand{\arraystretch}{1.15}
\begin{center}
\begin{tabular}{|c|c||c|c|c|}
\cline{3-5} %
\multicolumn{2}{c||}{ }  &  tree-weed    & tree-soil & tree-soil-weed   \\ %
\cline{2-5} %
\hline
\multirow{7}*{\rotatebox{90}{RE}} & LMM     & $16.4$          & $27.1$          & $6.80$ \\%
                                  & o-LMM      & $33.9$          & $50.0$          & $37.4$ \\%
                                  & FM      & $17.7$          & $16.4$          & $10.9$ \\%
                                  & o-FM      & $26.0$          & $40.7$          & $53.0$ \\%
                                    & NM    &  $16.3$         & $26.8$          & $2.13$ \\%
                                  & GBM     & $15.9$          & $15.2$          & $6.71$ \\%
                                  & PPNM    & $\best{3.07}$   & $\best{1.82}$   & $\best{1.21}$\\%
\hline %
\multirow{5}*{\rotatebox{90}{MSE}} & LMM   & $\second{12.5}$ & $2.78$          & $6.42$ \\%
                                    & FM    &  $13.5$         & $2.88$          & $8.15$ \\%
                                    & NM    &  ${12.6}$         & $\second{2.71}$          & $\second{5.80}$ \\%
                                    & GBM   & $\best{12.2}$   & ${2.86}$   & $6.39$ \\%
                                    & PPNM  & $13.0$          & $\best{2.57}$ & $\best{4.83}$  \\%
\hline %
\end{tabular}
\caption{Two- and three-endmember in-situ measurements. Abundance
MSE ($\times 10^{-2}$) and RE ($\times 10^{-4}$) for various
linear/nonlinear mixing models.\label{tab:RMSE_real}}
\end{center}
\end{table}

\begin{figure}[h!]
  \centering
  \includegraphics[width=0.75\columnwidth]{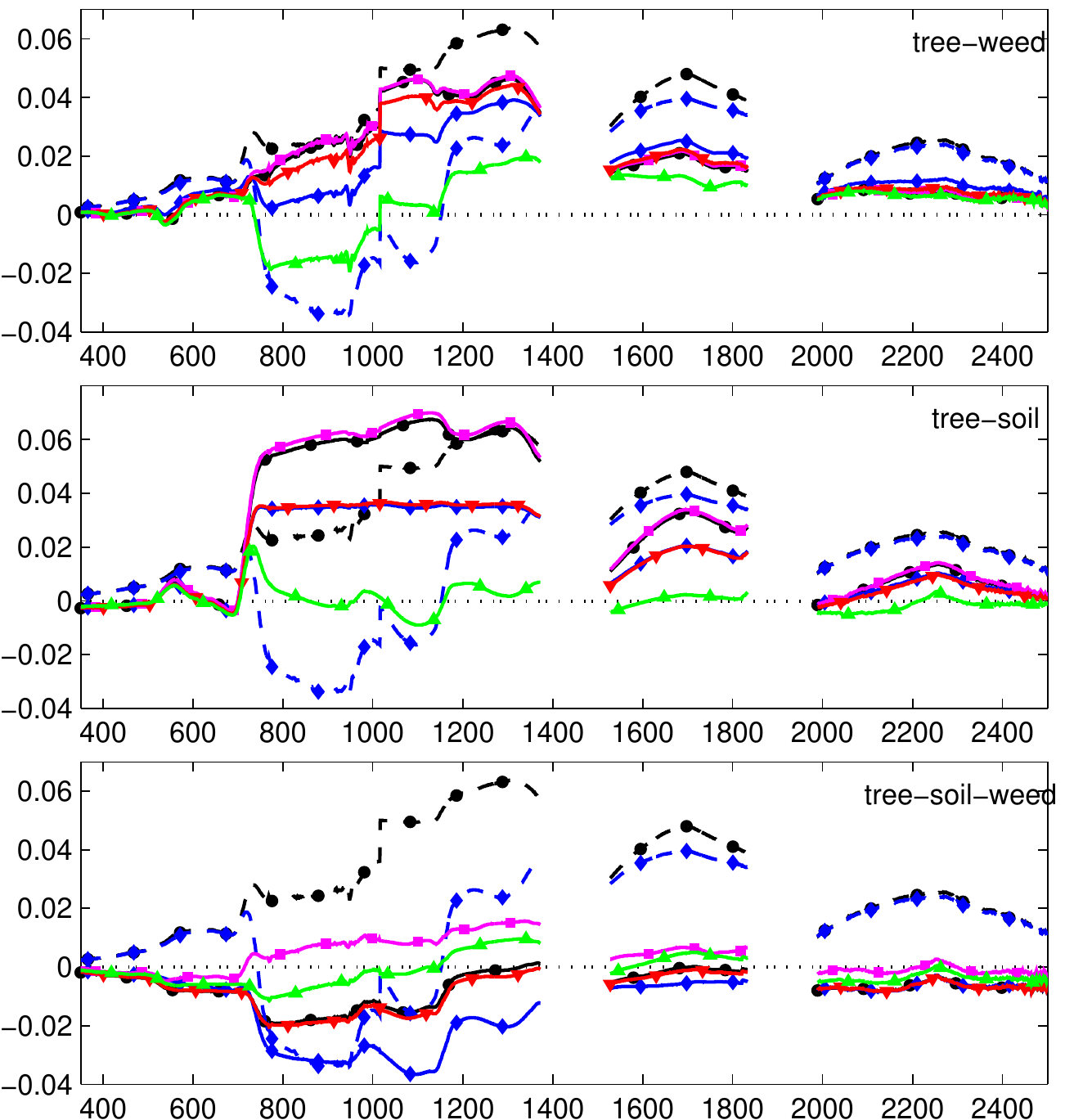}
  \caption{Two- and three-endmember in-situ measurements. Reconstruction difference $\textrm{RD}_{\ell}$ as a function of wavelength  for various linear/nonlinear mixing
models: LMM (black), oracle LMM (black, dashed line), FM (blue),
oracle FM (blue, dashed line), NM (magenta), GBM (red) and PPNM
(green).}
  \label{fig:in_situ_error_wavelength}
\end{figure}

The RDs obtained on the in-situ measurements are depicted in Fig.
\ref{fig:in_situ_error_wavelength}. Similarly to the previous
analyzed dataset, most of the nonlinear effects seem to occur in the
$700\textrm{nm}-1400\textrm{nm}$ spectral range, while being very
small in the visible range. From these plots, most of the mixing
model appear not sufficiently accurate to capture the nonlinearities
in the observed mixtures, except the PPNM.

\section{Discussion}
\label{sec:discussion}

The various datasets used during the experiments enable the
assessment of the performance of different unmixing models, and the
evaluation of the relevance of using nonlinear mixing models to
properly describe mixtures observed in vegetated areas. As the exact
per-pixel endmembers are known, the effects of endmember spectral
variability can be strongly reduced. Consequently, the simulated or
measured mixed pixels can be fully characterized by the abundances,
and the influence of the nonlinear mixing effects on the unmixing
accuracy could be evaluated. To qualitatively and quantitatively
evaluate the mixing models and corresponding unmixing algorithms,
general trends emerge from the results presented in
Section~\ref{sec:results}. These findings are reported in what
follows.

\subsection{Quantifying the amount of nonlinearity with o-LMM}

Since the endmember signatures as well as the abundance coefficients
are perfectly known for each pixel of the considered scenes, the
modeling error (i.e., the RE) obtained with the oracle-LMM could be
considered as the mis-modeling introduced by nonlinear mixing
effects. For all three data sets, a significant RE can be observed
with the oracle-LMM, demonstrating the presence of nonlinear mixing
effects, as already shown in \cite{Borel1994, Somers2009,
Tits2012igarss}, for example. In particular, the results reported in Table \ref{tab:RMSE_real} show that the in situ-measurements are submitted to highly nonlinear
effects. Conversely, from Table \ref{tab:synth_forest_RMSE}, the forest synthetic dataset seems to be less subjected to these nonlinear effects.
Overall, from the results reported in
the previous section, the mixed pixel signatures seem to be better
represented by nonlinear mixing models, and specifically PPNM and NM.
However, all nonlinear mixing models can not be advocated to better
describe mixed pixels than LMM, such as the GBM and NM for the
simulated orchard data (see Table \ref{tab:synth_orchard_RMSE}), and
the FM for the simulated forest data (see Table
\ref{tab:synth_forest_RMSE}) and the in-situ orchard data (see Table
\ref{tab:RMSE_real}). This shows that these nonlinear mixing models
do not necessarily better represent the mixed signatures.


\subsection{On the use of reconstruction error to assess a mixing model}
It is also important to note that a better modeling of the mixed
pixels does not necessarily result in a better estimation of the
abundances. For instance, PPNM, which has been shown to be the most
accurate to model nonlinearly mixed spectral signatures, sometimes
lead to less accuracy with respect to the abundance estimation when
compared to LMM, in particular for the three-endmember mixtures in the
simulated orchard data (see Table \ref{tab:synth_orchard_RMSE}) and
for the tree-weed mixtures in the in-situ data (see Table
\ref{tab:RMSE_real}). In the results of the simulated forest, the
same trend can be observed: in spite of increasing nonlinear mixing
effects, the REs remain almost constant for both the PPNM and the
NM, while the accuracy of the estimated abundances decreases (see
Table \ref{tab:synth_forest_RMSE}). As a consequence, the model
fitting error, widely used in the remote sensing literature to
monitor the performance of the unmixing algorithm, can not be used
as the unique figure-of-merit to evaluate the relevance of a given
mixing model.


\subsection{Mis-modeling with respect to wavelength}
All nonlinear mixing models considered in Section \ref{sec:models}
and used in the experiments reported in Section \ref{sec:results}
implicitly assume the same amount of nonlinearity for each
wavelength of the spectral domain. Indeed, they are basically
defined by cross-products between the endmember spectra, without
introducing any weighting functions that would depend on the
spectral bands. However, from the RDs depicted in Fig.'s
\ref{fig:synth_orchard_wavelength},
\ref{fig:synth_forest_wavelength} and
\ref{fig:in_situ_error_wavelength}, it clearly appears that the
mis-modeling is drastically subjected to the influence of the
wavelength. This corroborates the results of Somers \emph{et al.}
who also noticed similar behavior for the bilinear mixing model
\cite{Somers2014}. Most of the nonlinear models under test lead to
reconstructed mixtures with the same admissible accuracy as the LMM
in the visible range ($400\textrm{nm}-700\textrm{nm}$). Conversely,
a clear degradation of the modeling performance can be observed in
the $700\textrm{nm}-1400\textrm{nm}$ spectral range for most linear
and nonlinear models, except for the PPNM. In particular, the RDs
associated with the oracle-LMM demonstrate the important level of
nonlinearity in the near-infrared region. This finding has been widely observed in the literature \cite{Huete1985,Goel1988,Jacquemoud2000}.


\subsection{Dealing with the unmixing algorithm intrinsic limitations}
For both LMM and FM models, oracle measures of performance have been
computed since these models are fully described by the a priori
known abundance coefficients, explicitly considered as the spatial
distributions of the materials over the imaged pixels. However, for
the other nonlinear mixing models, unmixing algorithms need to be
used to infer all the parameters involved in the model specification
(e.g., abundances and nonlinearity parameters). Unfortunately, the
optimization problems to be solved, formulated in
\eqref{eq:criterion_LMM} and \eqref{eq:criterion_NLMM}, to recover
the abundance coefficients are not totally straightforward, mainly
due to the constraints and/or the nonlinearity. As a consequence,
the reliability of the obtained results, in terms of RE and
abundance MSE, should be carefully analyzed, indeed mitigated. More
precisely, part of the REs may consist of approximation errors
induced by the unmixing algorithms themselves, in particular when
these iterative algorithms converge toward a stationary point which
is not the global minimizer of the objective function. Consequently,
the abundance estimates may be biased since subjected to these
approximation errors. As a manifest example, one can consider the fitting performance of the GBM. By definition, this model generalizes both LMM and FM and, thus,
should provide at least similar RE to the lowest RE among those obtained with LMM and FM. However, this is not the case for
the orchard synthetic dataset, as already highlight in paragraph \ref{subsubsec:results_orchard} (see Table \ref{tab:synth_orchard_RMSE}). This is an
archetypal instance of the limitations of the GBM-based unmixing
algorithm.

\section{Conclusion}
\label{sec:conclusion}

This paper attempted to make a first step toward a full quantitative
assessment of linear and nonlinear mixing models to properly
described mixtures observed in hyperspectral images acquired over
vegetated areas. The conducted work exploited two kinds of
hyperspectral data, whose main advantages lies in the availability
of ground truth, that consists of the actual material signatures
(endmember spectra) and their corresponding spatial repartitions in
the pixels (abundance coefficients). The first set of hyperspectral
data consisted of physically-based simulated images, while the
second set of hyperspectral data came from real in-situ
measurements. Various linear and nonlinear mixing models were used
to analyze these data. They were evaluated in terms of spectral
mis-modeling (i.e., reconstruction error) and abundance estimation
accuracy. From the obtained results, it clearly appeared that the
polynomial post-nonlinear mixing model undeniably provided, by far,
the best reconstruction of the mixed pixels. It also persistently
led to admissible abundance estimates, regardless of the considered
scene. More generally, depending of the analyzed mixtures, the
Nascimento model, the Fan model or the polynomial post-nonlinear
model provided the most interesting results with respect to the
abundance estimates. However, it was worth noting that the results
presented in this work needed to be mitigated by the intrinsic
limitations of the resorted unmixing algorithms, that could induce
estimate biases. Finally, it is important to admit that the results  reported in this work are only valid for $2$- and $3$-endmember mixtures. Generalizing or extending these findings to more complex scenes would require further investigation.


\bibliographystyle{ieeetran}
\bibliography{D:/Dropbox/strings_all_ref,D:/Dropbox/all_ref}

\end{document}

%% file: com_notations.tex
\def\bfx{{\mathbf{x}}}



%% file: notations.tex
\newcommand{\Vpix}[1]{\mathbf{y}_{#1}}
\newcommand{\hatVpix}[1]{\hat{\mathbf{y}}_{#1}}

\newcommand{\pix}[2]{y_{#1,#2}}
\newcommand{\hatpix}[2]{\hat{y}_{#1,#2}}

\newcommand{\nbpix}{P}
\newcommand{\nopix}{p}

\newcommand{\nbband}{L}

\newcommand{\nbmat}{R}
\newcommand{\nomat}{r}

\newcommand{\MATmat}{{\mathbf M}}
\newcommand{\Vmat}[1]{{\mathbf m}_{#1}}
\newcommand{\mat}[2]{m_{#1,#2}}

\newcommand{\Vabond}[1]{{\mathbf a}_{#1}}
\newcommand{\hatVabond}[1]{\hat{\mathbf a}_{#1}}
\newcommand{\abond}[2]{a_{#1,#2}}

\newcommand{\Vnabond}[1]{{\mathbf b}_{#1}}
\newcommand{\hatVnabond}[1]{\hat{\mathbf b}_{#1}}
\newcommand{\nabond}[2]{b_{#1,#2}}

\newcommand{\Vnoise}[1]{\mathbf{n}_{#1}}

%% file: manuscript_arxiv.bbl
\begin{thebibliography}{10}
\providecommand{\url}[1]{#1}
\csname url@samestyle\endcsname
\providecommand{\newblock}{\relax}
\providecommand{\bibinfo}[2]{#2}
\providecommand{\BIBentrySTDinterwordspacing}{\spaceskip=0pt\relax}
\providecommand{\BIBentryALTinterwordstretchfactor}{4}
\providecommand{\BIBentryALTinterwordspacing}{\spaceskip=\fontdimen2\font plus
\BIBentryALTinterwordstretchfactor\fontdimen3\font minus
  \fontdimen4\font\relax}
\providecommand{\BIBforeignlanguage}[2]{{%
\expandafter\ifx\csname l@#1\endcsname\relax
\typeout{** WARNING: IEEEtran.bst: No hyphenation pattern has been}%
\typeout{** loaded for the language `#1'. Using the pattern for}%
\typeout{** the default language instead.}%
\else
\language=\csname l@#1\endcsname
\fi
#2}}
\providecommand{\BIBdecl}{\relax}
\BIBdecl

\bibitem{Bioucas2012jstars}
J.~M. Bioucas-Dias, A.~Plaza, N.~Dobigeon, M.~Parente, Q.~Du, P.~Gader, and
  J.~Chanussot, ``Hyperspectral unmixing overview: Geometrical, statistical,
  and sparse regression-based approaches,'' \emph{IEEE J. Sel. Topics Appl.
  Earth Observations and Remote Sens.}, vol.~5, no.~2, pp. 354--379, April
  2012.

\bibitem{Keshava2002}
N.~Keshava and J.~F. Mustard, ``Spectral unmixing,'' \emph{IEEE Signal Process.
  Mag.}, vol.~19, no.~1, pp. 44--57, Jan. 2002.

\bibitem{Somers2011}
B.~Somers, G.~P. Asner, L.~Tits, and P.~Coppin, ``Endmember variability in
  spectral mixture analysis: A review,'' \emph{Remote Sens. Environment}, vol.
  115, no.~7, pp. 1603--1616, July 2011.

\bibitem{Heinz2001}
D.~C. Heinz and {C. -I Chang}, ``Fully constrained least-squares linear
  spectral mixture analysis method for material quantification in hyperspectral
  imagery,'' \emph{IEEE Trans. Geosci. and Remote Sensing}, vol.~29, no.~3, pp.
  529--545, March 2001.

\bibitem{Theys2009}
C.~Theys, N.~Dobigeon, J.-Y. Tourneret, and H.~Lant\'eri, ``Linear unmixing of
  hyperspectral images using a scaled gradient method,'' in \emph{Proc. IEEE-SP
  Workshop Stat. and Signal Processing (SSP)}, Cardiff, UK, Aug. 2009, pp.
  729--732.

\bibitem{Heylen2011tgrs}
R.~Heylen, D.~Burazerovic, and P.~Scheunders, ``Fully constrained least squares
  spectral unmixing by simplex projection,'' \emph{IEEE Trans. Geosci. and
  Remote Sensing}, vol.~49, no.~11, pp. 4112--4122., Nov. 2011.

\bibitem{Dobigeon2008}
N.~Dobigeon, J.-Y. Tourneret, and {C.-I Chang}, ``Semi-supervised linear
  spectral unmixing using a hierarchical {B}ayesian model for hyperspectral
  imagery,'' \emph{IEEE Trans. Signal Process.}, vol.~56, no.~7, pp.
  2684--2695, July 2008.

\bibitem{Eches2010ip}
O.~Eches, N.~Dobigeon, C.~Mailhes, and J.-Y. Tourneret, ``{B}ayesian estimation
  of linear mixtures using the normal compositional model. {A}pplication to
  hyperspectral imagery,'' \emph{IEEE Trans. Image Process.}, vol.~19, no.~6,
  pp. 1403--1413, June 2010.

\bibitem{Eches2011icassp}
O.~Eches, N.~Dobigeon, J.-Y. Tourneret, and H.~Snoussi, ``Variational methods
  for spectral unmixing of hyperspectral unmixing,'' in \emph{Proc. IEEE Int.
  Conf. Acoust., Speech, and Signal Processing (ICASSP)}, Prague, Czech
  Republic, May 2011, pp. 957--960.

\bibitem{Dobigeon2014}
N.~Dobigeon, J.-Y. Tourneret, C.~Richard, J.~C.~M. Bermudez, S.~McLaughlin, and
  A.~O. Hero, ``Nonlinear unmixing of hyperspectral images: Models and
  algorithms,'' \emph{IEEE Signal Process. Mag.}, vol.~31, no.~1, pp. 89--94,
  Jan. 2014.

\bibitem{Hapke1981}
B.~W. Hapke, ``Bidirectional reflectance spectroscopy. {I}. {T}heory,''
  \emph{J. Geophys. Res.}, vol.~86, no.~B4, pp. 3039--3054, April 1981.

\bibitem{Guilfoyle2001}
K.~J. Guilfoyle, M.~L. Althouse, and C.-I. Chang, ``A quantitative and
  comparative analysis of linear and nonlinear spectral mixture models using
  radial basis function neural networks,'' \emph{IEEE Trans. Geosci. and Remote
  Sensing}, vol.~39, no.~8, pp. 2314--2318, Aug. 2001.

\bibitem{Nascimento2010}
J.~M.~P. Nascimento and J.~M. {Bioucas-Dias}, ``Unmixing hyperspectral intimate
  mixtures,'' in \emph{Proc. SPIE Image and Signal Processing for Remote
  Sensing XVI}, L.~Bruzzone, Ed., vol. 74830.\hskip 1em plus 0.5em minus
  0.4em\relax SPIE, Oct. 2010, p. 78300C.

\bibitem{Broadwater2009whispers}
J.~Broadwater and A.~Banerjee, ``A comparison of kernel functions for intimate
  mixture models,'' in \emph{Proc. IEEE GRSS Workshop Hyperspectral Image
  SIgnal Process.: Evolution in Remote Sens. (WHISPERS)}, Aug. 2009, pp. 1--4.

\bibitem{Broadwater2010whispers}
------, ``A generalized kernel for areal and intimate mixtures,'' in
  \emph{Proc. IEEE GRSS Workshop Hyperspectral Image SIgnal Process.: Evolution
  in Remote Sens. (WHISPERS)}, June 2010, pp. 1--4.

\bibitem{Broadwater2011whispers}
------, ``Mapping intimate mixtures using an adaptive kernel-based technique,''
  in \emph{Proc. IEEE GRSS Workshop Hyperspectral Image SIgnal Process.:
  Evolution in Remote Sens. (WHISPERS)}, Lisbon, Portugal, June 2011, pp. 1--4.

\bibitem{Close2012spie}
R.~Close, P.~Gader, A.~Zare, J.~Wilson, and D.~Dranishnikov, ``Endmember
  extraction using the physics-based multi-mixture pixel model,'' in
  \emph{Proc. SPIE Imaging Spectrometry XVII}, S.~S. Shen and P.~E. Lewis,
  Eds., vol. 8515.\hskip 1em plus 0.5em minus 0.4em\relax San Diego,
  California, USA: SPIE, Aug. 2012, pp. 85\,150L--14.

\bibitem{Close2012spieb}
R.~Close, P.~Gader, J.~Wilson, and A.~Zare, ``Using physics-based macroscopic
  and microscopic mixture models for hyperspectral pixel unmixing,'' in
  \emph{Proc. SPIE Algorithms and Technologies for Multispectral,
  Hyperspectral, and Ultraspectral Imagery XVIII}, S.~S. Shen and P.~E. Lewis,
  Eds., vol. 8390.\hskip 1em plus 0.5em minus 0.4em\relax Baltimore, Maryland,
  USA: SPIE, April 2012, pp. 83\,901L--83\,901L--13.

\bibitem{Heylen2014}
R.~Heylen and P.~Gader, ``Nonlinear spectral unmixing with a linear mixture of
  intimate mixtures model,'' \emph{IEEE Trans. Geosci. and Remote Sensing},
  vol.~11, no.~7, pp. 1195--1199, July 2014.

\bibitem{Borel1994}
C.~C. Borel and S.~A.~W. Gerstl, ``Nonlinear spectral mixing model for
  vegetative and soil surfaces,'' \emph{Remote Sens. Environment}, vol.~47,
  no.~3, pp. 403--416, 1994.

\bibitem{Ray1996}
T.~Ray and B.~Murray, ``Nonlinear spectral mixing in desert vegetation,''
  \emph{Remote Sens. Environment}, vol.~55, no.~1, pp. 59--64, 1996.

\bibitem{Zhang1998}
L.~Zhang, D.~Li, Q.~Tong, and L.~Zheng, ``Study of the spectral mixture model
  of soil and vegetation in poyang lake area, china,'' \emph{Remote Sens.
  Environment}, vol.~19, pp. 2077--2084, 1998.

\bibitem{Chen2006}
X.~Chen and L.~Vierling, ``Spectral mixture analyses of hyperspectral data
  acquired using a tethered balloon,'' \emph{Remote Sens. Environment}, vol.
  103, no.~3, pp. 338--350, Aug. 2006.

\bibitem{Fan2009}
W.~Fan, B.~Hu, J.~Miller, and M.~Li, ``Comparative study between a new
  nonlinear model and common linear model for analysing laboratory
  simulated-forest hyperspectral data,'' \emph{Int. J. Remote Sens.}, vol.~30,
  no.~11, pp. 2951--2962, June 2009.

\bibitem{Somers2009}
B.~Somers, K.~Cools, S.~Delalieux, J.~Stuckens, D.~V. der Zande, W.~W.
  Verstraeten, and P.~Coppin, ``Nonlinear hyperspectral mixture analysis for
  tree cover estimates in orchards,'' \emph{Remote Sens. Environment}, vol.
  113, pp. 1183--1193, Feb. 2009.

\bibitem{Tits2012igarss}
L.~Tits, W.~Delabastita, B.~Somers, J.~Farifteh, and P.~Coppin, ``First results
  of quantifying nonlinear mixing effects in heterogeneous forests: A modeling
  approach,'' in \emph{Proc. IEEE Int. Conf. Geosci. Remote Sens. (IGARSS)},
  2012, pp. 7185--7188.

\bibitem{Somers2014}
B.~Somers, L.~Tits, and P.~Coppin, ``Quantifying nonlinear spectral mixing in
  vegetated areas: computer simulation model validation and first results,''
  \emph{IEEE J. Sel. Topics Appl. Earth Observations and Remote Sens.}, 2014,
  to appear.

\bibitem{Huard2011whispers}
P.~Huard and R.~Marion, ``Study of non-linear mixing in hyperspectral imagery
  -- a first attempt in the laboratory,'' in \emph{Proc. IEEE GRSS Workshop
  Hyperspectral Image SIgnal Process.: Evolution in Remote Sens. (WHISPERS)},
  Lisbon, Portugal, June 2011, pp. 1--4.

\bibitem{Fontanilles2011}
G.~Fontanilles and X.~Briottet, ``A nonlinear unmixing method in the infrared
  domain,'' \emph{Appl. Opt.}, vol.~50, no.~20, pp. 3666--3677, July 2011.

\bibitem{Meganem2014}
I.~Meganem, P.~D\'eliot, X.~Briottet, Y.~Deville, and S.~Hosseini,
  ``Linear-quadratic mixing model for reflectances in urban environments,''
  \emph{IEEE Trans. Geosci. and Remote Sensing}, vol.~52, no.~1, pp. 544--558,
  Jan. 2014.

\bibitem{Altmann2011whispers}
Y.~Altmann, N.~Dobigeon, and J.-Y. Tourneret, ``Bilinear models for nonlinear
  unmixing of hyperspectral images,'' in \emph{Proc. IEEE GRSS Workshop
  Hyperspectral Image SIgnal Process.: Evolution in Remote Sens. (WHISPERS)},
  Lisbon, Portugal, June 2011, pp. 1--4.

\bibitem{Altmann2012ip}
Y.~Altmann, A.~Halimi, N.~Dobigeon, and J.-Y. Tourneret, ``Supervised nonlinear
  spectral unmixing using a post-nonlinear mixing model for hyperspectral
  imagery,'' \emph{IEEE Trans. Image Process.}, vol.~21, no.~6, pp. 3017--3025,
  June 2012.

\bibitem{Eches2014grsl}
O.~Eches and M.~Guillaume, ``A bilinear-bilinear non-negative matrix
  factorization method for hyperspectral unmixing,'' \emph{IEEE Geosci. and
  Remote Sensing Lett.}, vol.~11, no.~4, pp. 778--782, April 2014.

\bibitem{Stuckens2009}
J.~Stuckens, B.~Somers, S.~Delalieux, W.~W.~W. Verstraeten, and P.~Coppin,
  ``The impact of common assumptions on canopy radiative transfer simulations:
  a case study in citrus orchards,'' \emph{J. Quantitative Spectroscopy and
  Radiative Transfer}, vol. 110, no. 1--2, pp. 1--21, Jan. 2009.

\bibitem{VanderZande2010}
D.~V. der Zande, J.~S. W.~W. Verstraeten, B.~Muys, and P.~Coppin, ``Assessment
  of light dynamics in broadleaved forest canopies using terrestrial laser
  scanning,'' \emph{Remote Sensing}, vol.~2, no.~6, pp. 1564--1474, 2010.

\bibitem{Tits2012}
L.~Tits, W.~D. Keersmaecker, B.~Somers, G.~P. Asner, J.~Farifteh, and
  P.~Coppin, ``Hyperspectral shape-based unmixing to improve intra- and
  interclass variability for forest and agro-ecosystem monitoring,''
  \emph{ISPRS J. Photogrammetry and Remote Sensing}, vol.~74, pp. 163--174,
  Nov. 2012.

\bibitem{Nascimento2009spie}
J.~M.~P. Nascimento and J.~M. {Bioucas-Dias}, ``Nonlinear mixture model for
  hyperspectral unmixing,'' in \emph{Proc. SPIE Image and Signal Processing for
  Remote Sensing XV}, L.~Bruzzone, C.~Notarnicola, and F.~Posa, Eds., vol.
  7477, no.~1.\hskip 1em plus 0.5em minus 0.4em\relax SPIE, 2009, p. 74770I.

\bibitem{Halimi2011}
A.~Halimi, Y.~Altmann, N.~Dobigeon, and J.-Y. Tourneret, ``Nonlinear unmixing
  of hyperspectral images using a generalized bilinear model,'' \emph{IEEE
  Trans. Geosci. and Remote Sensing}, vol.~49, no.~11, pp. 4153--4162, Nov.
  2011.

\bibitem{Taleb1999}
A.~Taleb and C.~Jutten, ``Source separation in post-nonlinear mixtures,''
  \emph{IEEE Trans. Signal Process.}, vol.~47, no.~10, pp. 2807--2820, Oct.
  1999.

\bibitem{Altmann2013ip}
Y.~Altmann, N.~Dobigeon, and J.-Y. Tourneret, ``Nonlinearity detection in
  hyperspectral images using a polynomial post-nonlinear mixing model,''
  \emph{IEEE Trans. Image Process.}, vol.~22, no.~4, pp. 1267--1276, April
  2013.

\bibitem{Halimi2011igarss}
A.~Halimi, Y.~Altmann, N.~Dobigeon, and J.-Y. Tourneret, ``Unmixing
  hyperspectral images using the generalized bilinear model,'' in \emph{Proc.
  IEEE Int. Conf. Geosci. Remote Sens. (IGARSS)}, Vancouver, Canada, July 2011,
  pp. 1886--1889.

\bibitem{VanderZande2008}
D.~Van~der Zande, ``{Mathematical Modeling of 3D Canopy Structure in Forest
  Stands Using Ground-Based Lidar},'' Ph.D. dissertation, Katholieke
  Universiteit Leuven, Leuven, Belgium, 2008.

\bibitem{Pharr2004}
M.~Pharr and G.~Humphreys, \emph{{Physically Based Rendering: From Theory to
  Implementation}}, T.~Cox, Ed.\hskip 1em plus 0.5em minus 0.4em\relax San
  Fransisco: Morgan Kaufmann Pub, 2004.

\bibitem{Weber1995}
J.~Weber and J.~Penn, ``Creation and rendering of realistic trees,'' in
  \emph{Proc. SIGGRAPH Annu. Conf. Comput. Graph. Interactive Techn.}, 1995,
  pp. 119--128.

\bibitem{FAO1998}
FAO, ``{FAO} world reference base for soil resources,'' Food and Agriculture
  Organisation of the United Nations, Rome, World Soil Resource Report~84,
  1998.

\bibitem{Somers2010a}
B.~Somers, V.~Gysels, W.~Verstraeten, S.~Delalieux, and P.~Coppin, ``Modelling
  moisture-induced soil reflectance changes in cultivated sandy soils: a case
  study in citrus orchards,'' \emph{European J. Soil Sci.}, vol.~61, no.~6, pp.
  1091--1105, 2010.

\bibitem{Hosgood1995}
B.~Hosgood, S.~Jacquemoud, G.~Andreoli, J.~Verdebout, A.~Pedrini, and
  G.~Schmuck, ``Leaf {O}ptical {P}roperties {EX}periment 93 ({LOPEX93}),''
  Joint Research Centre / Institute for Remote Sensing Applications Unit for
  Advanced Techniques, Ispra, Italy, Report EUR 16095 EN, 1995.

\bibitem{Huete1985}
A.~R. Huete, R.~D. Jackson, and D.~F. Post, ``Spectral response of a plant
  canopy with different soil backgrounds,'' \emph{Remote Sens. Environment},
  vol.~17, no.~1, pp. 37--53, Feb. 1985.

\bibitem{Goel1988}
N.~S. Goel, ``Models of vegetation canopy reflectance and their use in
  estimation of biophysical parameters from reflectance data,'' \emph{Remote
  Sens. Rev}, vol.~4, no.~1, pp. 1--212, 1988.

\bibitem{Jacquemoud2000}
S.~Jacquemoud, C.~Bacour, H.~Poilv\'e, and J.-P. Frangi, ``Comparisons of four
  radiative transfer models to simulate plant canopies reflectance: Direct and
  inverse mode,'' \emph{Remote Sens. Environment}, vol.~74, no.~3, pp.
  471--481, Dec. 2000.

\end{thebibliography}
